# The Pressure-Stabilized Polymorph of Indium Triiodide

Danrui Ni,[a] Haozhe Wang,[b] Xianghan Xu,[a] Weiwei Xie,[b] and Robert J. Cava[a]*

[a]Department of Chemistry, Princeton University, Princeton, NJ 08544, USA
[b]Department of Chemistry, Michigan State University, East Lansing, MI 48824, USA

*Corresponding author: Robert J. Cava (rcava@princeton.edu)

**Abstract**

A layered rhombohedral polymorph of indium (III) triiodide is synthesized at high pressure and temperature. The unit cell symmetry and approximate dimensions are determined by single crystal X-ray diffraction. Its *R*-3 crystal structure, with *a* = 7.217 Å, and *c* = 20.476 Å, is refined by the Rietveld method on powder X-ray diffraction data. The crystal structure is based on $InI_6$ octahedra sharing edges to form honeycomb lattice layers, though with considerable stacking variations. Different from ambient pressure $InI_3$, which has a monoclinic molecular structure and a light-yellow color, high pressure $InI_3$ is layered and has an orange color. The band gaps of both the monoclinic and rhombohedral variants of $InI_3$ are estimated from diffuse reflectance measurements.

*Keywords:* Triiodide; Polymorphism; High pressure synthesis; Layered structure.

**Introduction**

High pressure synthesis and the influence of high-pressure treatment have attracted researchers' attention in recent years, as the application of pressure can lead to significant changes in reaction equilibria and affect both the structures and physical properties of compounds[1–3]. High-pressure high-temperature synthetic techniques may introduce phase transitions and lead to different polymorphs[4–6], or result in new formulas or structures[7–9] that cannot be stabilized using traditional preparation methods at ambient pressure.

 Metal trihalides often crystallize in crystal structures consisting of 1D-chains or 2D-honeycomb layers [10–12]. Polymorphism and phase transition behavior, as well as the structure-related properties of different phases, are aspects of their study [13–16], and pressure-induced polymorphism has been reported in several transition-metal-based systems[17–19]. In this context, some main-group-metal halide systems such as the indium iodide system[20] have also been reported to have relatively complex polymorphic behavior. Based on information in the Inorganic Crystal Structure Database, indium triiodide ($InI_3$) is suggested to have an aluminum triiodide type monoclinic structure with $In_2I_6$ dimers at ambient pressure[21]. Another $InI_3$ polymorph, mentioned in the literature as early as the 1940s[22] as a low-temperature phase, was suggested to be light-sensitive and isostructural to the low-temperature phase of CrCl$_3$[20], but no modern structural refinement results were reported. In this report, we describe an *R*-3 polymorph of InI$_3$ with a honeycomb layered structure that is stabilized by high pressure synthesis. The structure was determined by powder X-ray diffraction (PXRD) via Rietveld refinement, with the space group symmetry and cell parameters qualitatively determined by single crystal X-day diffraction (SCXRD). Stacking faults occur in significant amounts in high pressure InI$_3$, which smear out the diffraction peaks (shown below), making single crystal structure refinements unreliable. Finally, some light absorption and physical properties are characterized.

**Experiments**

The ambient-pressure phase of monoclinic InI$_3$ was purchased (from Alfa Aesar, anhydrous, 99.999%) and used as the starting material for the high-pressure synthesis. To do that synthesis, the ground powder of the starting material was loaded into a boron nitride crucible in an air-free atmosphere, inserted into a pyrophyllite cube assembly and pressed to 6 GPa using a cubic multi-anvil system (Rockland Research Corporation). The sample was annealed at 500 °C for 3 hours with the temperature measured by an internal thermocouple. The system was then cooled before decompression. The products obtained were reddish in color, from which small single crystals were taken for SCXRD characterization. The ground powder of the post-reaction sample shows an orange-red color, which is different from the light-yellow color of the monoclinic InI$_3$ starting material.

Due to the air-sensitivity and hygroscopicity of the materials, the PXRD patterns used for structural characterization were collected using a Rigaku Miniflex II diffractometer located inside a nitrogen-filled glove box for both the starting material and the products, with Cu Kα radiation (λ= 1.5406 Å) employed. Le Bail fitting was performed on the starting material pattern to confirm its uniformity (via TOPAS software), while a full Rietveld refinement was performed via GSAS II on the high-pressure-synthesis product diffraction pattern for structural analysis.

To determine the space group symmetry and cell parameters, a crystal of HP-InI$_3$ with dimensions 0.216 × 0.105 × 0.084 mm$^3$ was picked up, mounted on a nylon loop with paratone oil, and characterized using a XtalLAB Synergy, Dualflex, Hypix single crystal X-ray diffractometer with an Oxford Cryosystems low-temperature device, operating at *T* = 300(2) K. The diffraction data were collected using ω scans with Mo Kα radiation (λ= 0.71073 Å, micro-focus sealed X-ray tube, 50 kV, 1 mA). The total number of runs and images was based on the intensity collection strategy calculation from the program CrysAlisPro 1.171.42.79a (Rigaku OD, 2022). Cell parameters were retrieved and refined based on 5708 reflections, 50.6% of those observed. Data reduction was performed with correction for Lorentz polarization. A numerical absorption correction based on gaussian integration, as implemented in SCALE3 ABSPACK was applied. The SHELXTL Software Package[23,24], was used to determine the space group as *R*-3, with *Z* = 6, and an approximate crystal structure.

Magnetization and heat capacity were measured using a Quantum Design PPMS (Dynacool), equipped with a vibrating sample magnetometer (VSM) option. The temperature-dependent magnetization (*M*) was measured in an applied magnetic field (*H*) of 1000 Oe. UV-Vis diffuse reflectance spectra are collected using an Agilent Cary 5000 spectrometer with Agilent Internal DRA-2500 diffuse reflectance accessory on powder samples. Samples were diluted with dry MgO to 50% w/w, while dry MgO is used as the reflectance standard. The band gap values are calculated using Tauc plots, while reflectance data were converted to absorption using the Kubelka-Munk function. Transfers of samples to the measurement apparatus were performed very rapidly to protect the sample from hydrolysis. Both AP- and HP-InI$_3$ are air-sensitive and hygroscopic. They absorb moisture in air and change into a white color in several minutes.

**Results and Discussion**

The purity of the ambient pressure (AP-InI$_3$) starting material was confirmed by Le Bail fitting, as was its *P*21/*c* monoclinic symmetry (Figure 1A and 1B), consistent with the reported structure [21]. Different from the light-yellow colored AP-InI$_3$ powder, the sample after high pressure reaction (HP-InI$_3$, shown in Figure 1C) appears to be orange after grinding (Figure 1B and 1D insets). Rietveld refinement of the PXRD data of HP-InI$_3$, collected at 300 K in an air-free environment (Figure 1D and Table 1), suggests that it has an *R*-3 space group (#148) with lattice parameters *a* = 7.2166(6) Å and a *c* = 20.4762(12) Å. Due to a significant amount of disorder and stacking errors, the refinement gives 75.3% of the total indium content located at the honeycomb site (6*c* Wyckoff site), and 24.7% at a site in the honeycomb hole (the 3*a* site). The rhombohedral unit cell contains three layers of honeycomb lattice material (Figure 1C), formed by edge-sharing InI$_6$ octahedra. The space group symmetry is confirmed by the single crystal diffraction data shown in Figure 2, especially in the (hk0) and (hk1) reciprocal lattice planes, although some ambiguous scattering may be observed due to the disorder in the system. (The streaking in the (h0l) and (0kl) reciprocal lattice planes is an indication of the disorder in the system.)

In HP-InI$_3$, the indium has a relatively symmetric octahedral coordination, with the In-I bond lengths ranging from 2.895 to 2.913 Å in the InI$_6$ octahedra. When comparing these two polymorphs of InI$_3$, it is revealed that the In atoms in AP-InI, which display a tetrahedral coordination, bond to iodine with distances at around 2.64 Å (the bonds to the iodines forming the unshared edges of the dimers) and 2.84 Å (the bonds forming the shared edges of the dimers)[21]. These In-I bond lengths are generally shorter than the ones in HP-InI$_3$, which suggests a stronger In-I bonding in the In$_2$I$_6$ dimer of AP-InI$_3$. However, based on the structures presented in Figure 1, AP-InI$_3$ is closer to being a molecular solid, in other words a much weaker interaction between different dimer units is observed. This leads to the significantly lower density of AP-InI$_3$ (4.72 g/cm$^3$) compared to HP-InI$_3$ (5.35 g/cm$^3$), and thus makes the R-3 polymorph a more favored phase under pressure. This structural type, based on stacked honeycomb layers, is relatively common in metal trihalides[14].

The unit cell and color of the HP-InI$_3$ sample match the description of the "low-temperature polymorph" of InI$_3$ [20,25]. Although the latter paper is inclined to assign an *R*3 symmetry to that polymorph based on a Hamilton significance test, we have not found evidence for the absence of an inversion center when determining the structure of HP-InI$_3$ and thus attribute *R*-3 to the phase in this report. Previous reports also indicated that this phase is light-sensitive (while ours is not), and that it would transform to monoclinic AP-InI$_3$ under light radiation, or by heating with a transition temperature ranging from 23 to 69 °C. Our observations confirm that HP-InI$_3$ is metastable, and that the samples may undergo a partial phase transition due to a delicate stimulus from the environment. However, it was also observed that with good protection, an HP-InI$_3$ sample was unchanged in either color or structure under ordinary room light at room temperature in a N$_2$-protected glove box. This could be because a longer time, stronger light intensity, or higher energy radiation is needed to trigger HP-InI$_3$ to transform into the AP-phase than is needed for the low temperature form of InI$_3$. Another possible explanation is that an external factor (for example a tiny amount of moisture) might be needed to accelerate the phase transition. (This phenomenon has been observed in some halide systems, one instance being that CsPbI$_3$ was reported to catalytically transfer from a perovskite phase to a non-perovskite phase in the presence of moisture[26].)

Diffuse reflectance measurements were conducted on both AP-InI$_3$ and HP-InI$_3$. Due to the metastability of the HP-phase, however, the measured sample might have partially transformed to the ambient pressure phase during the measurement and thus two transitions can be observed in the HP-InI$_3$ sample reflectance curve (Figure 3). According to the comparison of the two reflectance curves, and the correlation with their sample colors, we attribute the absorption between 500 to 600 nm to the HP-InI$_3$ phase, and a second absorption at around 400 nm to the AP-phase. The band gaps were calculated using Tauc plots (Figure 3 inset) based on the equation[27]:

$$(\alpha h\nu)^n = A (h\nu - E_g) \qquad (1)$$

where *A* is a constant and $\alpha$ is the absorption coefficient (cm$^{-1}$). (*n* is 2 for direct transitions, and 0.5 for indirect transitions.) For indirect transitions, the values of the band gaps are calculated to be 2.71 eV for AP-InI$_3$, and 2.01 eV for HP-InI$_3$. DFT calculations obtained from the Open Quantum Materials Database (OQMD) provide calculated density of states for both *R*-3 structure InI$_3$ and *P*2$_1$/*c* structure InI$_3$ [28,29]. Based on that information, *R*-3 InI$_3$ has a calculated band gap of around 1.8 eV, while *P*2$_1$/*c* InI$_3$ is calculated to have a band gap of 2.2 eV. The larger band gap value of the monoclinic phase compared to the rhombohedral phase matches the experimental results of the diffuse reflectance measurements.

The magnetic susceptibility (*M*/*H*) was measured versus increasing temperature on an HP-InI$_3$ powder sample under 1000 Oe. As is shown in the inset of Figure 4, temperature-independent diamagnetic behavior is revealed, which is in line with the expectation of the electron configuration of In(III). Heat capacity measurements were also conducted from 2 to 200 K on dense HP-InI$_3$ sample pieces. As presented in Figure 4's main panel, no sharp transitions are observed, consistent with the magnetization behavior of the sample. Some small slope variation can be observed on the curve, which looks similar to what might happen for a second-order phase transition. A transition at around 93 K has been mentioned in previous reports[20]. These phase transitions may be due to a minor structural transformation, suggesting that more complex polymorphic behavior may be found in the indium iodine system than is presently known. Thus it may be of future research interest to determine the low temperature crystal structure of HP-InI$_3$.

**Conclusion**

A high-pressure stabilized InI$_3$ polymorph was obtained by a high-pressure high-temperature synthetic method. Its rhombohedral structure and *R*-3 unit cell have been determined by SCXRD and Rietveld refinement of PXRD data. The orange color and honeycomb layered structure distinguish this HP-phase from monoclinic AP-InI$_3$. Diffuse reflectance measurements are also conducted, from which the band gap values are estimated to be 2.71 eV for AP-InI$_3$ and 2.01 eV for HP-InI$_3$, for indirect transitions. Magnetization and heat capacity have also been characterized, and suggest that further study of the polymorphic behavior of the indium iodine system may be of future interest. Compared with the description in previous reports, high pressure plays a crucial role in synthesis and forces the structure towards layered honeycomb InI$_3$, which further confirms the potential of high pressure synthesis in exploring and developing new solid-state materials.

**Acknowledgements**

This work was funded in large part by the Gordon and Betty Moore foundation, EPiQS initiative, grant GBMF-9066. The SCXRD work performed in Weiwei Xie's group was supported by the U.S. DOE-BES under Contract DE-SC0022156.

**References**

[1] Badding JV. High-pressure synthesis, characterization, and tuning of solid state materials. Annu Rev Mater Sci 1998;28:631–58. https://doi.org/10.1146/annurev.matsci.28.1.631.

[2] Polvani DA, Meng JF, Chandra Shekar NV, Sharp J, Badding JV. Large Improvement in Thermoelectric Properties in Pressure-Tuned p-Type $Sb_{1.5}Bi_{0.5}Te_3$. Chem Mater 2001;13:2068–71. https://doi.org/10.1021/cm000888q.

[3] Lorenz B, Chu CW. High Pressure Effects on Superconductivity. In: Narlikar AV, editor. Frontiers in Superconducting Materials, Berlin, Heidelberg: Springer; 2005, p. 459–97. https://doi.org/10.1007/3-540-27294-1_12.

[4] Shirako Y, Wang X, Tsujimoto Y, Tanaka K, Guo Y, Matsushita Y, et al. Synthesis, Crystal Structure, and Electronic Properties of High-Pressure $PdF_2$-Type Oxides $MO_2$ (M = Ru, Rh, Os, Ir, Pt). Inorg Chem 2014;53:11616–25. https://doi.org/10.1021/ic501770g.

[5] von Rohr FO, Ji H, Cevallos FA, Gao T, Ong NP, Cava RJ. High-Pressure Synthesis and Characterization of β-GeSe—A Six-Membered-Ring Semiconductor in an Uncommon Boat Conformation. J Am Chem Soc 2017;139:2771–7. https://doi.org/10.1021/jacs.6b12828.

[6] Nawa K, Imai Y, Yamaji Y, Fujihara H, Yamada W, Takahashi R, et al. Strongly Electron-Correlated Semimetal $RuI_3$ with a Layered Honeycomb Structure. J Phys Soc Jpn 2021;90:123703. https://doi.org/10.7566/JPSJ.90.123703.

[7] Ni D, Guo S, Yang ZS, Powderly KM, Cava RJ. $Pb_4S_3I_2$–A high-pressure phase in the $PbS-PbI_2$ system. Solid State Sciences 2019;91:49–53. https://doi.org/10.1016/j.solidstatesciences.2019.03.012.

[8] Ni D, Guo S, Powderly KM, Zhong R, Cava RJ. A high-pressure phase with a non-centrosymmetric crystal structure in the $PbSe–PbBr_2$ system. Journal of Solid State Chemistry 2019;280:120982. https://doi.org/10.1016/j.jssc.2019.120982.

[9] Ni D, Gui X, Han B, Wang H, Xie W, Ong NP, et al. The non-centrosymmetric layered compounds $IrTe_2I$ and $RhTe_2I$. Dalton Trans 2022;51:8688–94. https://doi.org/10.1039/D2DT01224C.

[10] Hillebrecht H, Ludwig T, Thiele G. About Trihalides with $TiI_3$ Chain Structure: Proof of Pair Forming of Cations in β-$RuCl_3$ and $RuBr_3$ by Temperature Dependent Single Crystal X-ray Analyses. Zeitschrift Für Anorganische Und Allgemeine Chemie 2004;630:2199–204. https://doi.org/10.1002/zaac.200400106.

[11] Banerjee A, Yan J, Knolle J, Bridges CA, Stone MB, Lumsden MD, et al. Neutron scattering in the proximate quantum spin liquid α-$RuCl_3$. Science 2017;356:1055–9. https://doi.org/10.1126/science.aah6015.

[12] Kong T, Stolze K, Timmons EI, Tao J, Ni D, Guo S, et al. $VI_3$—a New Layered Ferromagnetic Semiconductor. Advanced Materials 2019;31:1808074. https://doi.org/10.1002/adma.201808074.

[13] Angelkort J, Schönleber A, van Smaalen S. Low- and high-temperature crystal structures of $TiI_3$. Journal of Solid State Chemistry 2009;182:525–31. https://doi.org/10.1016/j.jssc.2008.11.028.

[14] McGuire MA. Crystal and Magnetic Structures in Layered, Transition Metal Dihalides and Trihalides. Crystals 2017;7:121. https://doi.org/10.3390/cryst7050121.

[15] Chamorro JR, McQueen TM, Tran TT. Chemistry of Quantum Spin Liquids. Chem Rev 2021;121:2898–934. https://doi.org/10.1021/acs.chemrev.0c00641.

[16] Ni D, Devlin KP, Cheng G, Gui X, Xie W, Yao N, et al. The honeycomb and hyperhoneycomb polymorphs of IrI$_3$. Journal of Solid State Chemistry 2022;312:123240. https://doi.org/10.1016/j.jssc.2022.123240.

[17] Imai Y, Nawa K, Shimizu Y, Yamada W, Fujihara H, Aoyama T, et al. Zigzag magnetic order in the Kitaev spin-liquid candidate material RuBr$_3$ with a honeycomb lattice. Phys Rev B 2022;105:L041112. https://doi.org/10.1103/PhysRevB.105.L041112.

[18] Ni D, Gui X, Powderly KM, Cava RJ. Honeycomb-Structure RuI$_3$, A New Quantum Material Related to α-RuCl3. Advanced Materials 2022;34:2106831. https://doi.org/10.1002/adma.202106831.

[19] Ni D, Mudiyanselage RSD, Xu X, Mun J, Zhu Y, Xie W, et al. Layered polymorph of titanium triiodide. Phys Rev Mater 2022;6:124001. https://doi.org/10.1103/PhysRevMaterials.6.124001.

[20] Fedorov PP, Popov AI, Simoneaux RL. Indium iodides. Russ Chem Rev 2017;86:240. https://doi.org/10.1070/RCR4609.

[21] Forrester JD, Zalkin A, Templeton DH. Crystal and Molecular Structure of Indium(III) Iodide (In$_2$I$_6$). Inorg Chem 1964;3:63–7. https://doi.org/10.1021/ic50011a013.

[22] Ensslin F, Dreyer H. Über einige Salze des Indiums. I. Zeitschrift für anorganische und allgemeine Chemie 1942;249:119–32. https://doi.org/10.1002/zaac.19422490111.

[23] Sheldrick GM. Crystal structure refinement with *SHELXL*. Acta Crystallographica Section C Structural Chemistry 2015;71:3–8. https://doi.org/10.1107/S2053229614024218.

[24] Sheldrick GM. SHELXT – Integrated space-group and crystal-structure determination. Acta Cryst A 2015;71:3–8. https://doi.org/10.1107/S2053273314026370.

[25] Kniep R, Blees P. Red α-Indium(III) Iodide. Angewandte Chemie International Edition in English 1984;23:799–800. https://doi.org/10.1002/anie.198407991.

[26] Straus DB, Guo S, Cava RJ. Kinetically Stable Single Crystals of Perovskite-Phase CsPbI$_3$. J Am Chem Soc 2019;141:11435–9. https://doi.org/10.1021/jacs.9b06055.

[27] Tauc J. Optical properties and electronic structure of amorphous Ge and Si. Materials Research Bulletin 1968;3:37–46. https://doi.org/10.1016/0025-5408(68)90023-8.

[28] Saal JE, Kirklin S, Aykol M, Meredig B, Wolverton C. Materials Design and Discovery with High-Throughput Density Functional Theory: The Open Quantum Materials Database (OQMD). JOM 2013;65:1501–9. https://doi.org/10.1007/s11837-013-0755-4.

[29] Kirklin S, Saal JE, Meredig B, Thompson A, Doak JW, Aykol M, et al. The Open Quantum Materials Database (OQMD): assessing the accuracy of DFT formation energies. Npj Comput Mater 2015;1:1–15. https://doi.org/10.1038/npjcompumats.2015.10.

**Table 1.** Structural parameters and standardized crystallographic positions of HP-InI$_3$, from the Rietveld refinement of laboratory PXRD data collected at 300 K. wR = 14.212%, GOF = 2.86, reduced $\chi^2$ = 8.20. Due to the multiplicity differences, the 6*c* site accounts for 75.3% of the In total while the 3*a* site accounts for 24.7%.

| Refined Formula | InI$_3$ |
|---|---|
| F.W. (g/mol) | 495.53 |
| Symmetry | Trigonal |
| Space group; Z | R-3; 6 |
| a(Å) | 7.2166(6) |
| c(Å) | 20.4762(12) |
| V (Å$^3$) | 923.53(13) |
| Density (g/cm$^3$) | 5.35 |

| Atom | Wyck. | Occ. | x | y | z | $U_{iso}$ equiv. [Å$^2$] | Multi. |
|---|---|---|---|---|---|---|---|
| I1 | 18*f* | 1 | 0.3534(6) | 0.0047(6) | 0.08372(13) | 0.0037(11) | 18 |
| In2 | 6*c* | 0.753 | 0.66670 | 0.33330 | -0.0010(8) | 0.018(4) | 6 |
| In3 | 3*a* | 0.494 | 0.00000 | 0.00000 | 0.00000 | 0.043(12) | 3 |

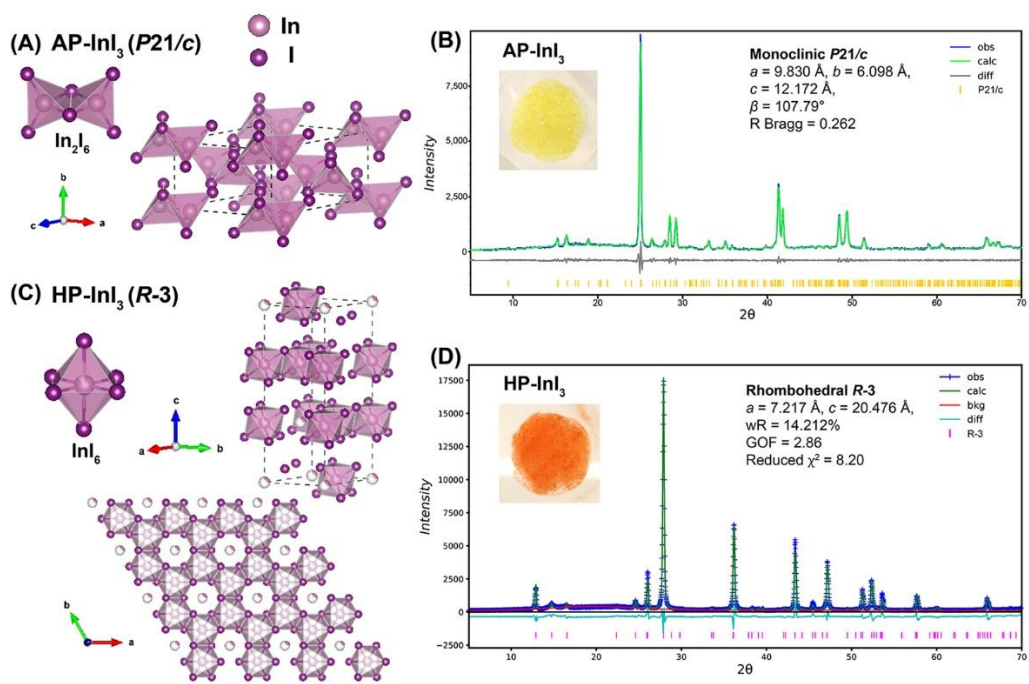

**Figure 1.** (A) Crystal structure of monoclinic AP-InI$_3$ with In$_2$I$_6$ tetragonal dimer; (B) Le Bail fitting of PXRD pattern of AP-InI$_3$, with a photo of powder sample shown as inset; (C) Crystal structure of *R*-3 HP-InI$_3$, with a view through *c*-direction to reveal the in-plane honeycomb lattice, as well as the InI$_6$ coordination octahedron; (D) Rietveld refinement of HP-InI$_3$ PXRD pattern, with a photo of powder sample shown as the inset. The reflection positions are labeled with colored sticks in the powder diffraction patterns.

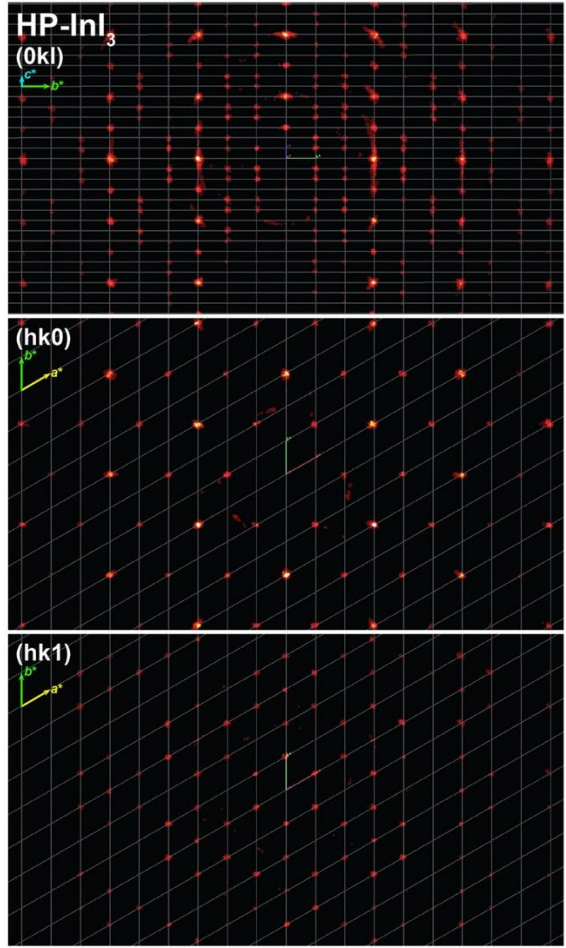

**Figure 2.** The 0kl (top) hk0 (middle) and hk1 (bottom) reciprocal lattice planes of single crystal HP-InI$_3$ at 300 K.

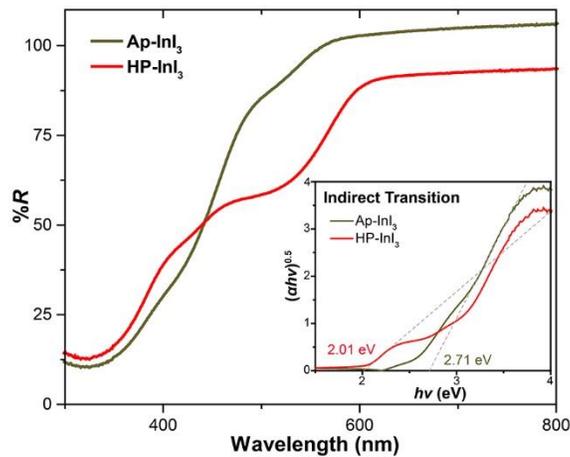

**Figure 3**. Diffuse reflectance spectra of AP-InI$_3$ and HP-InI$_3$ samples. The inset presents the Tauc plots of indirect transitions to estimate the band gap values.

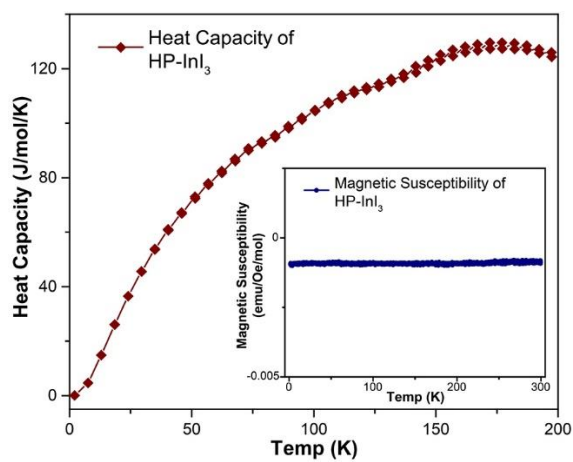

**Figure 4.** Magnetic susceptibility (inset) and heat capacity (main panel) data of HP-InI$_3$, plotted versus temperature.